# Experimental study of MnCl$_3$(C$_{12}$H$_8$N$_2$) - an S = 2 Heisenberg antiferromagnetic chain

E. Čižmár,[a,d,*] J.-H. Park,[a] S. J. Gamble,[a] B. H. Ward,[b] D. R. Talham,[b] J. van Tol,[c] L.-C. Brunel,[c] M. Orendáč,[d] A. Feher,[d] J. Šebek,[e] M. W. Meisel[a]

[a]*Department of Physics and Center for Condensed Matter Sciences, University of Florida, Gainesville FL 32611-8440, USA*

[b]*Department of Chemistry, University of Florida, Gainesville FL32611-7200, USA*

[c]*NHMFL, Florida State University, Tallahassee FL 32310-3706, USA*

[d]*Institute of Physics, P.J. Šafárik University, Park Angelinum 9, 04154 Košice, Slovakia*

[e]*Institute of Physics, AS CR, Na Slovance 2, 18221 Prague 8, Czech Republic*

**Abstract**

The thermodynamic and magnetic properties of MnCl$_3$(C$_{12}$H$_8$N$_2$), have been studied down to 2 K. The material is an S = 2 antiferromagnetic linear chain that undergoes long-range ordering to a canted antiferromagnetic state at 23 K. The comparison of magnetic data from polycrystalline samples with the results of specific heat measurements using a pressed pellet of powder reveals the importance of the length of the chains in the sample in relation to the magnetic correlation length.

Since Haldane's conjecture that antiferromagnetic (AFM) quantum Heisenberg chains with integer spin S have an energy gap between the non-magnetic ground state and the first excited state [1], significant experimental and theoretical progress has been made in understanding these systems [2]. Experimental evidence of the Haldane gap has been observed in several S = 1 materials [3] and only in one S = 2 material, MnCl$_3$(C$_{10}$H$_8$N$_2$) [4]. The magnetic properties of a similar material, MnCl$_3$(C$_{12}$H$_8$N$_2$), were previously studied between 100 – 300 K [5] and down to 2 K in a preliminary study [6]. Other measurements below 100 K were briefly mentioned in [7], but the data have never been published. Due to a twinning of the crystals, the exact structure of MnCl$_3$(C$_{12}$H$_8$N$_2$) has not been determined [8]. Nevertheless, the structural data confirm the presence of linear chains of Mn$^{3+}$ ions surrounded by C$_{12}$H$_8$N$_2$ groups and bridged by Cl atoms, similar to MnCl$_3$(C$_{10}$H$_8$N$_2$) [9].

Using a commercial SQUID magnetometer, the temperature dependence of the magnetic susceptibility of MnCl$_3$(C$_{12}$H$_8$N$_2$) has been studied on a polycrystalline sample (Fig. 1). The difference between the zero-field cooled and field cooled data suggests long-range ordering below 23 K. Since the magnetic response did not exhibit any frequency dependent behavior up to 6 kHz, a spin glass state is not a plausible explanation. In addition, the field dependence of the magnetization at 5 K, 10 K and 20 K shows no saturation up to 7 T, but a spin-flop transition is observed at ≈ 50 mT. These data are consistent with the long-range canted antiferromagnetic ordering at 23 K. Canting of the spins in an antiferromagnetic chain is usually only a matter of few degrees and produces a small net magnetic moment $\mu_{net}$ in one direction, thereby forming the weak ferromagnetism [10]. From the difference between the zero-field cooled and field cooled susceptibility, we can estimate a lower limit of $\mu_{net} \approx 0.05\mu_B$. The presence of the canting might arise from the tilted orientation of adjacent octahedra formed by the ligands around the Mn$^{3+}$ ions.

[*] Corresponding author. Tel.: 1-352-392-9147; fax: 1-352-392-7709; e-mail: cizmar@phys.ufl.edu.

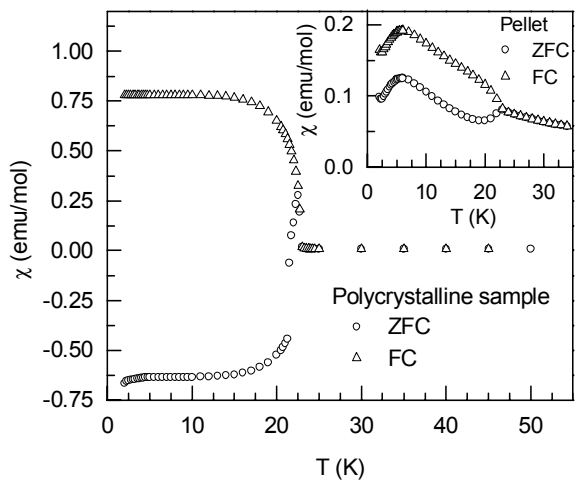

Figure 1 Susceptibility ($\chi$) versus temperature in 10 mT is shown when a polycrystalline sample is zero-field cooled (ZFC) or field cooled (FC). The inset shows the data from a pellet specimen.

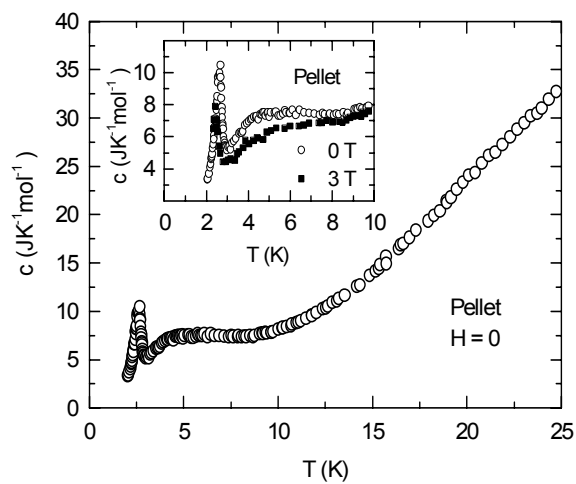

Figure 2 Temperature dependence of the specific heat c of pellet sample is given for magnetic fields up to 3 T.

The specific heat of a pressed pellet sample has been studied using an adiabatic pulse method on a $^4$He-cryostat in the temperature range of 2 – 26 K and in magnetic fields up to 3 T. The experimental data plotted in Fig. 2 show long-range ordering at 2.75 K and a Schottky-like anomaly at 5 K. The maximum value of the $\lambda$-anomaly at 2.75 K decreases with applied magnetic field and its position shifts to lower temperatures, suggesting the presence of antiferromagnetic exchange coupling. No anomaly has been observed at 23 K, where the susceptibility data show long-range ordering. A plausible explanation for the lack of evidence for ordering in the specific heat data is that the sample was prepared by applying pressure to form a pellet from a polycrystalline powder, and this process likely causes the chains to break into small fragments [11]. Consequently, the characteristic structural length of the chain is small when compared to the magnetic correlation length $\xi$, where $\xi = 61$ spin sites in an S = 2 chain [12]. In contrast, the magnetic correlation length is only 6 spin sites in an S = 1 chain [13]. For chain lengths smaller than $\xi$, the magnetic properties cannot be treated as purely one-dimensional.

In summary, the magnetic measurements of the nascent sample (polycrystalline powder) suggest the presence of long-range ordered canted antiferromagnetism below 23 K. The resulting net magnetic moment is $\approx 0.05\mu_B$. The specific heat and magnetic measurements on the pellet sample show the importance of structural chain length in a real sample in comparison with magnetic correlation length. In addition, high-field, high-frequency ESR data indicate an increased concentration of $Mn^{2+}$ impurity ions in the pellet sample. The magnitude of the canting angle can be determined using antiferromagnetic resonance [10].

**Acknowledgements**

Supported, in part, by the NSF through INT-0089140, DMR-0113714, DGE-0209410, and DMR-0084173, which supports the NHMFL. Additional support is gratefully acknowledged from the Slovak Grant Agency through VEGA 1/7473/20. We thank G.E. Granroth for early contributions to this work.